\documentclass[5p, preprint]{elsarticle}
\biboptions{sort&compress}
\usepackage{lineno,isotope,mathrsfs}
\usepackage{amsmath,bbold,amssymb,epsfig,feynmp,color,ifpdf}
\usepackage{slashed,nicefrac,exscale,multirow,times,txfonts}
\usepackage{epstopdf}
\usepackage[colorlinks,linkcolor=blue,anchorcolor=black,urlcolor=blue,citecolor=blue,CJKbookmarks=True]{hyperref}

\begin{document}
%%%frontmatter%%%%%%%%%%%%%%%%%%%%%%%%%%%%
%%%%%%%%%%%%%%%%%%%%%%%%%%%%%%%%%%%%%%%%%%
\begin{frontmatter}
\title{Electric dipole polarizability in neutron-rich Sn isotopes as a probe of nuclear isovector properties }
\author[a]{Z. Z. Li}
\author[a]{Y. F. Niu}\ead{niuyf@lzu.edu.cn}
\author[a]{W. H. Long}
\address[a]{School of Nuclear Science and Technology, Lanzhou University, Lanzhou 730000, China}
\begin{abstract}
The determination of nuclear symmetry energy, and in particular, its density dependence, is a long-standing problem for nuclear physics community. Previous studies have found that the product of electric dipole polarizability $\alpha_D$ and symmetry energy at saturation density $J$ has a strong linear correlation with $L$, the slope parameter of symmetry energy. However, current uncertainty of $J$ hinders the precise constraint on $L$. We investigate the correlations between electric dipole polarizability $\alpha_D$ (or times symmetry energy at saturation density $J$) in Sn isotopes and the slope parameter of symmetry energy $L$ using the quasiparticle random-phase approximation based on Skyrme Hartree-Fock-Bogoliubov. A strong and model-independent linear correlation between $\alpha_D$ and $L$ is found in neutron-rich Sn isotopes where pygmy dipole resonance (PDR) gives a considerable contribution to $\alpha_D$, attributed to the pairing correlations playing important roles through PDR. This newly discovered linear correlation would help one to constrain $L$ and neutron-skin thickness $\Delta R_\textnormal{np}$ stiffly if $\alpha_D$ is measured with high resolution in neutron-rich nuclei. Besides, a linear correlation between $\alpha_D J$ in a nucleus around $\beta$-stability  line and $\alpha_D$ in a neutron-rich nucleus can be used to assess $\alpha_D$ in neutron-rich nuclei.
\end{abstract}
\end{frontmatter}
%%%end frontmatter%%%%%%%%%%%%%%%%%%%%%%%%%%
%%%%%%%%%%%%%%%%%%%%%%%%%%%%%%%%%%%%%%%%%%
\section{Introduction}
The determination of nuclear equation of state (EoS) at high density is a challenge for both experimental and theoretical nuclear physics \cite{B.A.Li_2008_PhysRep_EOS, Oertel_2017_RMP_EoS},  which is crucial for constraining current theoretical models \cite{Dutra_2012_PRC_NM, Dutra_2014_PRC_NM} and understanding many phenomena in astrophysics \cite{Lattimer_2001_APJ_NS, Z.W.Liu_2018_PRC_EsymNS}. The biggest uncertainty of EoS comes from its isovector parts, which are governed by the nuclear symmetry energy $\mathcal S(\rho)$. The symmetry energy can be expanded as a function of $\varepsilon=(\rho-\rho_0)/3\rho_0$ by
\begin{equation}
\mathcal S (\rho) = J + L\varepsilon + \dfrac{1}{2}K_\textnormal{sym}\varepsilon^2 + ...
\end{equation}
where $J=\mathcal S(\rho_0)$ is the symmetry energy at saturation density $\rho_0$, while $L = 3\rho_0 \Big(\dfrac{\partial \mathcal S}{\partial \rho} \Big) \Big|_{\rho=\rho_0}$ and $K_\textnormal{sym} = 9 \rho_0^2 \Big(\dfrac{\partial^2 \mathcal S}{\partial \rho^2} \Big)\Big|_{\rho=\rho_0}$ correspond to the slope and curvature parameters at saturation density, respectively.

The slope parameter of symmetry energy $L$ determines the behavior of symmetry energy at high density, however, it varies a lot in different nuclear models. Constraints on $L$ can be obtained from heavy-ion collisions \cite{B.A.Li_2008_PhysRep_EOS, M.B.Tsang_2009_PRL_ESymHIC}, properties of neutron stars \cite{Lattimer_2001_APJ_NS, Lattimer_2016_PhysRep_EOS}, and nuclear properties of ground state and excited states of finite nuclei \cite{Roca-Maza_2018_PPNP_EoS}. For example, it is revealed that $L$ is proportional to the neutron-skin thickness $\Delta R_{np}$ by droplet model \cite{Myers_1980_NPA_Rnp, Warda_2009_PRC_Rnp}, which is further conformed by many microscopic models \cite{B.A.Brown_2000_PRL_Rnp, L.W.Chen_2005_PRC_Rnp}. However, the obstacle in the measurements of neutron radius hinders the access to high-resolution neutron skin data. As an alternative, charge radii difference $\Delta R_c$ between mirror nuclei is proposed as another possible way to constrain $L$ \cite{N.Wang_2013_PRC_DRc, B.A.Brown_2017_PRL_DRc, J.J.Yang_2018_PRC_DRc}, which also faces difficulties in the measurements of charge radius in proton-rich nucleus.

The electric dipole $(E1)$ excitation in nucleus is mainly composed of the giant dipole resonance (GDR), which is formed by the relative dipole oscillation between neutrons and protons, thus reflecting asymmetry information in nuclear EoS. The electric dipole polarizability $\alpha_D$, being proportional to the inverse energy-weighted sum rule of $E1$ excitation, can be served as a possible probe for nuclear isovector properties. Theoretically, (quasiparticle) random-phase approximation [(Q)RPA] approach  is widely used to describe small oscillations of nucleus, such as $E1$ excitations. The self-consistent (Q)RPA models have been developed based on Skyrme density functionals \cite{Colo_2013_CPC_RPA, Terasaki_2005_PRC_QRPA, Khan_2000_PLB_QRPA}, Gogny density functionals \cite{Giambrone_2003_NPA_GHF-QRPA, Martini_2011_PRC_GQRPA}, and relativistic density functionals \cite{Paar_2003_PRC_RQRPA, Paar_2007_RPP, Ring_2001_NPA_RRPA, Niksic_2002_PRC_RRPA}. Global properties of GDR, such as centroid energies and electric dipole polarizabilities, can be well described within this approximation.

Based on these self-consistent (Q)RPA models, the correlations between electric dipole polarizability $\alpha_D$ and other nuclear isovector properties have been investigated in recent years. Calculations performed by RPA model based on Skyrme density functionals SV-min series \cite{Klupfel_2009_PRC_SHF} and relativistic density functionals RMF-$\delta$-t series in $^{208}$Pb suggested a strong linear correlation between $\alpha_D$ and neutron-skin thickness $\Delta R_\textnormal{np}$ \cite{Reinhard_2010_PRC_alphaD}. However, when one combines the results from a host of different nuclear density functionals, this linear correlation is not universal anymore \cite{Piekarewicz_2012_PRC_Rn208Pb}. Starting from droplet model, and further supported by RPA calculations based on many different Skyrme and relativistic density functionals in $^{208}$Pb, the product of dipole polarizability and symmetry energy at saturation density $\alpha_DJ$ was suggested to be much better correlated with neutron-skin thickness and symmetry energy slope parameter $L$ than $\alpha_D$ alone is \cite{Roca-Maza_2013_PRC_alphaD}. Based on this correlation, $L=43\pm(6)_\textnormal{expt}\pm(8)_\textnormal{theor}\pm(12)_\textnormal{est}$ MeV was given by using the experimental $\alpha_D$ value in $^{208}$Pb \cite{Roca-Maza_2013_PRC_alphaD}, and the intervals $J=30-35$ MeV and $L=20-66$ MeV were further obtained by combining the measured polarizabilities in $^{68}$Ni, $^{120}$Sn and $^{208}$Pb \cite{Roca-Maza_2015_PRC_alphaD}.  Below saturation density, $\alpha_D$ in $^{208}$Pb was also found to be sensitive to both the symmetry energy $\mathcal S(\rho_c)$ and slope parameter $L(\rho_c)$ at the subsaturation cross density $\rho_c=0.11$fm$^{-3}$ \cite{Z.Zhang_2014_PRC_alphaD}. Since $\mathcal S (\rho_c)$ is well constrained, $L(\rho_c)$ can be strongly constrained from experimental $\alpha_D$ in $^{208}$Pb \cite{Z.Zhang_2014_PRC_alphaD}. At $\rho_r=\rho_0/3$, another linear correlation was built between $\alpha_D^{-1}$ and $\mathcal S(\rho_r)$ \cite{Z.Zhang_2015_PRC_alphaD}. Besides, $\alpha_D$ between two different nuclei \cite{Hashimoto_2015_PRC_alphaD}, as well as $\alpha_D J$ between two different nuclei \cite{Roca-Maza_2015_PRC_alphaD}, were also shown to have good linear correlations.

In recent years, the electric dipole polarizabilities $\alpha_D$ in $^{208}$Pb \cite{Tamii_2011_PRL_alphaD}, $^{48}$Ca \cite{Birkhan_2017_PRL_alphaD}, and stable Sn isotopes \cite{Hashimoto_2015_PRC_alphaD, Bassauer_2020_PRC_dipole, Bassauer_2020_PLB_alphaD} were measured with high resolution via polarized proton inelastic scattering at extreme forward angles \cite{Neumann-Cosel_2019_EPJA_scattering}. For unstable nucleus $^{68}$Ni, $\alpha_D$ was also extracted by Coulomb excitation in inverse kinematics \cite{Rossi_2013_PRL_alphaD}. However, there are problems when one uses these high-resolution dipole polarizability data to constrain isovector properties: the constraints on $L$ or $\Delta R_\textnormal{np}$ is either with big uncertainties due to the uncertainty of $J$ or in model-dependent ways. One way to solve the problem and constrain $L$ stiffly is to find a direct and model-independent correlation between $\alpha_D$ and $L$. Although the previous studies have shown that the model-independent linear correlation only exists between $\alpha_D J$ and $L$, it was only limited to stable nuclei or nuclei near $\beta$-stability line. It is well known that exotic phenomena will present when approaching to nuclei far from $\beta$-stability line, such as novel shell structures \cite{Wienholtz_2013_Nature_N32, J.Liu_2020_PLB_N3234, J.J.Li_2019_PLB_Si48, Z.Z.Li_2019_CPC_PSS, Grasso_2014_PRC_N3234}, new types of excitations \cite{Savran_2013_PPNP_PDR, Paar_2007_RPP, Aumann_2019_EPJA_PDR}, and so on. For $E1$ excitations, the pygmy dipole resonance (PDR) appears in neutron-rich nuclei \cite{Savran_2013_PPNP_PDR, Paar_2007_RPP, Aumann_2019_EPJA_PDR}, which would cause different characteristics of $E1$ excitations compared to the ones around $\beta$-stability line, and further affect $\alpha_D$. So an interesting question is if the linear correlation between $\alpha_DJ$ and $L$ observed in stable nuclei still holds and new correlations would appear in neutron-rich nuclei.

Therefore, in our study we will explore the correlations between $\alpha_D$  and nuclear isovector properties such as slope parameter $L$ and neutron-skin thickness $\Delta R_{np}$ in even-even Sn isotopes from neutron-deficient $^{100}$Sn to neutron-rich $^{164}$Sn. The calculations are performed by QRPA based on Skyrme Hartree-Fock-Bogoliubov (HFB) model, in which the spherical symmetries are imposed. The linear correlations are evaluated by a least-square regression analysis. Based on the newly discovered correlations, constraints on $L$ and neutron-skin thickness will be discussed.

\section{Theoretical Framework}
We carry out a self-consistent HFB$+$QRPA calculation of $E1$ strength using 19 Skyrme functionals: SIII, SIV, SV, SVI \cite{Beiner_1975_NPA_SHF}, SLy230a, SLy230b, SLy4, SLy5, SLy8 \cite{Chabanat_1997_NPA_SHF, Chabanat_1998_NPA_SHF}, SAMi \cite{Roca-Maza_2012_PRC_SHF}, SAMi-J30, SAMi-J31, SAMi-J32, SAMi-J33 \cite{Roca-Maza_2013_PRC_SAMiJ}, SGI, SGII \cite{VanGiai_1981_PLB_SHF}, SkM \cite{Krivine_1980_NPA_SHF}, SkM* \cite{Bartel_1982_NPA_SHF}, Ska \cite{Kohler_1976_NPA_SHF}. The detailed formulas of QRPA on top of HFB can be found in Ref. \cite{Terasaki_2005_PRC_QRPA}. The density-dependent zero-range surface pairing force is implemented in the particle-particle channel,
\begin{equation}
V_{pp}(\pmb r_1,\pmb r_2) = V_0 \Big[1- \dfrac{\rho(\pmb r)}{\rho_0} \Big] \delta(\pmb r_1 -\pmb r_2)	
\end{equation}
where $\pmb r=(\pmb r_1 +\pmb r_2)/2$, and $\rho_0=0.16$fm$^{-3}$ is the nuclear saturation density, while $V_0$ is adjusted by fitting neutron pairing gaps of $^{116\sim130}$Sn according to the five-point formula \cite{Bender_2000_EPJA_PGaps}. The electric dipole polarizability $\alpha_D$ is given by
\begin{equation}
\alpha_D = \dfrac{8\pi e^2}{9} m_{-1}, \quad m_{-1}= \sum\limits_{\nu} \dfrac{\big| \langle \psi_\nu| F_{1\mu}^\textnormal{(IV)}| \psi_0\rangle \big|^2}{E_\nu}
\end{equation}
where $\psi_\nu$ and $E_\nu$ are the eigenstates and eigenvalues of QRPA equations, and $\psi_0$ is the ground state. $m_{-1}$ is the inverse energy-weighted sum rule (EWSR), which is calculated using the isovector dipole operator
\begin{equation}
F_{1\mu}^{\textnormal{(IV)}} = \dfrac{N}{A} \sum\limits_{p=1}^Z r_pY_{1\mu} - \dfrac{Z}{A} \sum\limits_{n=1}^N r_nY_{1\mu}
\end{equation}
where $A$, $N$, $Z$ denote mass number, neutron number, proton number, and $Y_{1\mu}$ are the spherical harmonics. In our calculations, the quasiparticle energy cutoff $E_\textnormal{cut}$ is set as 90 MeV and the total angular momentum cutoff of quasiparticle $j_\textnormal{max}$ is set as $21/2$ to ensure the convergence of numerical results.

\section{Results and Discussions}

\subsection{Correlations between $\alpha_D$ and nuclear isovector properties}\label{Sec-3.1}

\begin{table}[b]
	\caption{Pearson's coefficient $r$ between the product of dipole polarizability and saturated symmetry energy $\alpha_D J$ and the slope parameter of symmetry energy $L$ in Sn isotopes, as well as the corresponding slope $k$ of the regression line ($\alpha_D J$ as a function of $L$), calculated by QRPA based on HFB with 19 Skyrme density functionals.}
	\renewcommand{\arraystretch}{1.3}\setlength{\tabcolsep}{0.1em}\setlength{\tabcolsep}{0.4em} \label{tab-1}
	\centering
	\begin{tabular}{clllllll} \hline\hline
		Nucleus        & $^{100}$Sn & $^{110}$Sn & $^{120}$Sn & $^{130}$Sn & $^{140}$Sn & $^{150}$Sn & $^{160}$Sn    \\ \hline
		$r$     & 0.965      & 0.966      & 0.974      & 0.961      & 0.966      & 0.940      & 0.937      \\ \hline
		$k$ (fm$^3$)     & 0.844      & 1.066      & 1.383      & 1.543      & 2.272      & 2.880      & 3.541      \\ \hline \hline
	\end{tabular}
\end{table}

\begin{figure*}[t]
	\centering
	\includegraphics[width=1.0\textwidth]{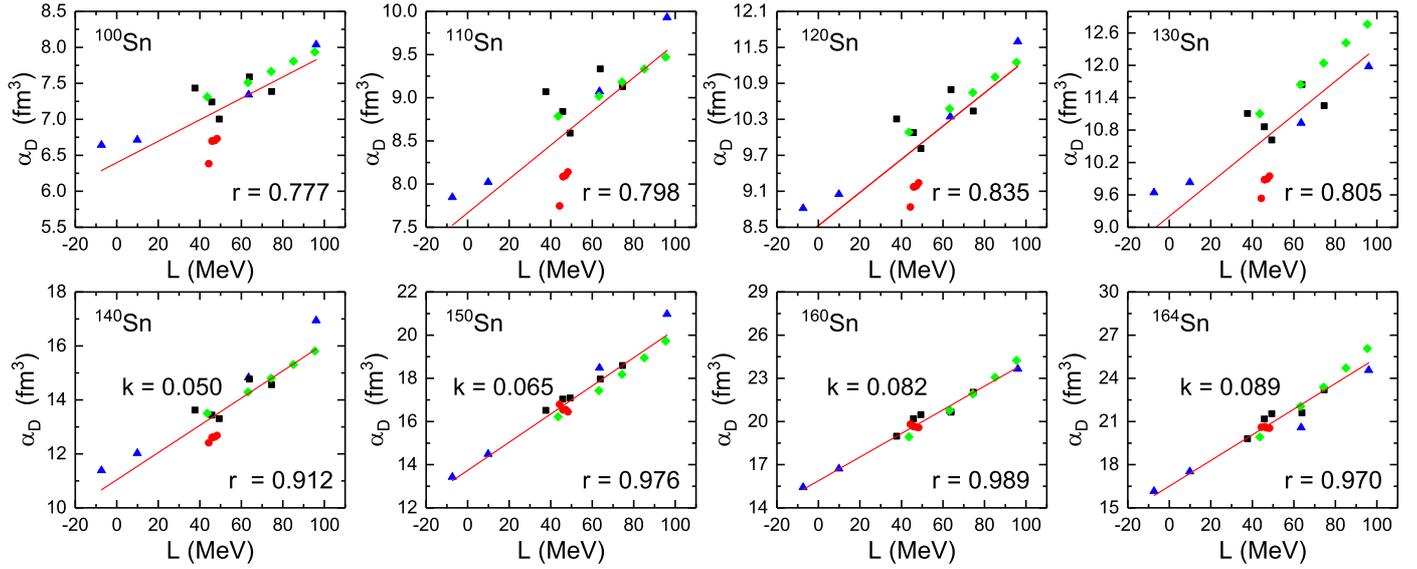}
	\caption{(Color online) Plots for dipole polarizability $\alpha_D$ against slope parameter of symmetry energy $L$ in Sn isotopes calculated by QRPA based on HFB with 19 Skyrme density functionals: SIII, SIV, SV, SVI (blue up triangles); SLy230a, SLy230b, SLy4, SLy5, SLy8 (red circles);  SAMi, SAMi-J30, SAMi-J31, SAMi-J32, SAMi-J33 (green diamonds); SGI, SGII, SkM, SkM*, Ska (black squares). A regression line (red solid line) is obtained by a least-square linear fit of the calculated $\alpha_D$ as a function of $L$. $r$ is Pearson's coefficient and $k$ (fm$^3/$MeV) is the slope of the regression line.} \label{fig-1}
\end{figure*}

\begin{figure}[t]
	\centering
	\includegraphics[width=0.45\textwidth]{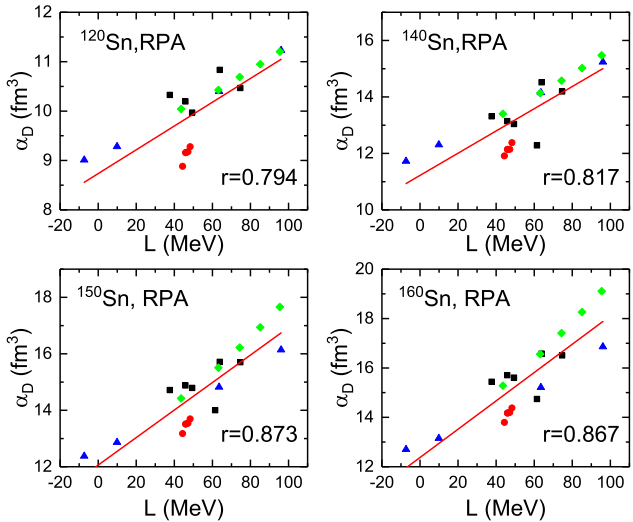}
	\caption{(Color online) The same as Fig. \ref{fig-1} but for $^{120,140,150,160}$Sn without the pairing correlations.} \label{fig-2}
\end{figure}

First of all, we study if the previously discovered linear correlation between $\alpha_D J$ and $L$ holds in the whole tin isotopes from neutron-deficient ones to neutron-rich ones. So in Tab. \ref{tab-1}, Pearson correlation coefficients (or Pearson's coefficients) $r$ between $\alpha_D J$ and $L$ in even-even Sn isotopes from $^{100}$Sn to $^{160}$Sn , as well as the corresponding slopes $k$ of the regression lines are shown based on the HFB+QRPA calculations using 19 Skyrme density functionals. Pearson's coefficient $r$ is a statistic that measures linear correlation between two variables, which is defined by the covariance of two variables divided by the product of their standard deviations. A value of $|r|=1$ means that the two observables are fully linearly correlated while $r=0$ are totally uncorrelated. From Tab. \ref{tab-1}, one can see the Pearson's coefficients $r$ in the whole Sn isotopes are all above $0.9$, showing strong linear correlations between $\alpha_D J$ and $L$. So it further proofs this linear correlation is a universal one which exists not only in stable nuclei as revealed in previous studies \cite{Roca-Maza_2013_PRC_alphaD}  but also in neutron-deficient and neutron-rich nuclei. The corresponding slope $k$ of the regression line shows a clear increase trend with the increase of neutron number. The larger $k$ value means a more rapid increase of $\alpha_D J$ as a function of $L$, which gives a smaller range of $L$ under the same uncertainty of $\alpha_D J$. So the slope $k$ of the regression line is an important quantity to select good candidate nuclei as probes of nuclear isovector properties, which will be discussed in details in Sec. \ref{Sec-3.2}.

Although the above correlation is universal, it cannot provide a stiff constraint on the slope parameter of symmetry energy $L$ due to the uncertainty in the symmetry energy $J$. For example, by adopting  $J=31\pm 2$ MeV, \textit{Roca-Maza et al.} obtained $L=43\pm(6)_\textnormal{expt}\pm(8)_\textnormal{theor}\pm(12)_\textnormal{est}$ MeV, where the uncertainty $\pm 12$ MeV comes from the uncertainty of $J$ \cite{Roca-Maza_2013_PRC_alphaD}. So it would be better to find a direct correlation between $\alpha_D$ and $L$. Previous studies have shown that $L$ and $\alpha_D$ have a good linear correlation within some specific parameter family \cite{Reinhard_2010_PRC_alphaD}, however, by including different parameter families, this correlation becomes bad, for example, in $^{208}$Pb the Pearson's coefficient r was given as $r=0.62$ \cite{Roca-Maza_2013_PRC_alphaD} and $r=0.77$ \cite{Piekarewicz_2012_PRC_Rn208Pb}. Here we recheck the correlation between the dipole polarizability $\alpha_D$ and the slope parameter $L$ of symmetry energy for the whole tin isotopes from neutron-deficient ones to neutron-rich ones, as shown in Fig. \ref{fig-1}, to see if the previous conclusions still hold. In stable nucleus $^{120}$Sn, for some specific Skyrme parameter family, such as SAMi (green diamonds) or SIII-SVI (up blue triangles), one can observe a good linear correlation, in agreement with Ref. \cite{Reinhard_2010_PRC_alphaD}. However, when one includes more different Skyrme parameter sets, the linear correlation becomes poor, and the Pearson coefficient $r$ is around 0.8, again in agreement with the case in $^{208}$Pb \cite{Roca-Maza_2013_PRC_alphaD,Piekarewicz_2012_PRC_Rn208Pb}. Similar situations still exist in nuclei not far from the stability line such as $^{100,110,130}$Sn.

However, the cases become totally different in the neutron-rich nuclei. The coefficients are above $0.9$ for the isotopes with mass number $A\ge140$, which present strong correlations between $\alpha_D$ and $L$ in the neutron-rich Sn isotopes. After $A\ge146$, the correlation between $\alpha_D$  and $L$ is even better than the one  between $\alpha_D J$ and $L$. We stress here the assessments are carried out by different Skyrme functional families. For the neutron-rich nuclei of $A\ge140$ with a clear linear correlation, we further give the slopes $k$ of the regression lines. It is seen that $k$ becomes larger with the increase of neutron number, which implies that the more neutron rich the nucleus is, the better probe it can be served as for nuclear isovector properties, seeing detailed discussions in Sec. \ref{Sec-3.2}.

\begin{figure}[t]
	\centering
	\includegraphics[width=0.42\textwidth]{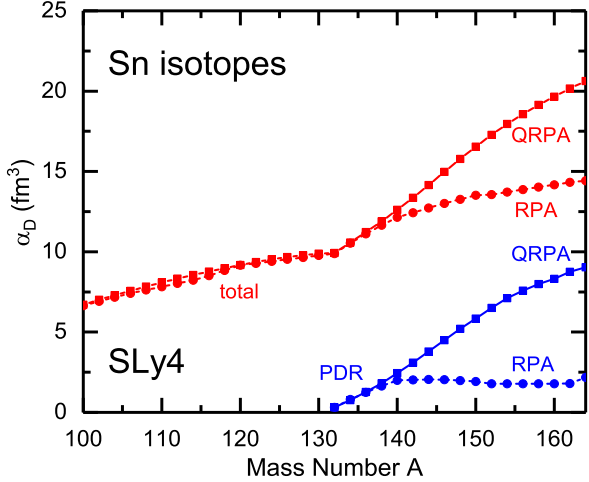}
	\caption{(Color online) The dipole polarizabilities as functions of mass number $A$ in even-even Sn isotopes calculated by QRPA (square line) and RPA (circle line) using Skyrme functional SLy4. The total dipole polarizabilities (red) and the contributions from PDR (blue) are shown respectively.} \label{fig-3}
\end{figure}
\begin{figure}[t]
	\centering
	\includegraphics[width=0.45\textwidth]{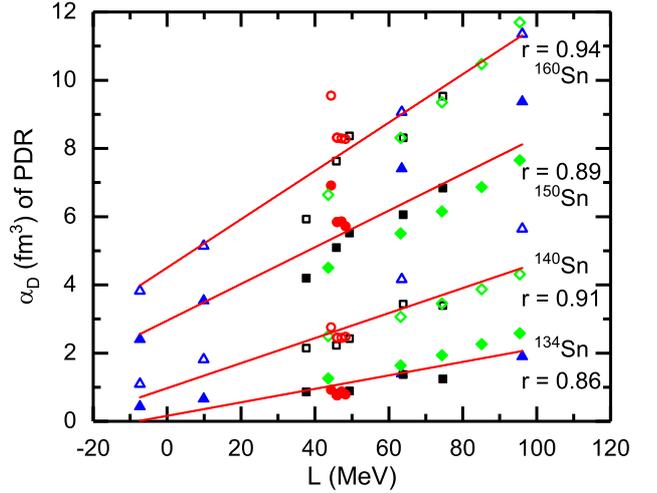}
	\caption{(Color online) Plots for dipole polarizability contributed by PDR against slope parameter of symmetry energy in $^{134,140,150,160}$Sn isotopes calculated by QRPA based on HFB with 19 Skyrme density functionals: SIII, SIV, SV, SVI (blue up triangles); SLy230a, SLy230b, SLy4, SLy5, SLy8 (red circles);  SAMi, SAMi-J30, SAMi-J31, SAMi-J32, SAMi-J33 (green diamonds); SGI, SGII, SkM, SkM*, Ska (black squares). A linear fit is done for each nucleus (red solid line) with a corresponding Pearson's coefficient $r$.}\label{fig-4}
\end{figure}

\begin{figure}[b]
	\centering
	\includegraphics[width=0.47\textwidth]{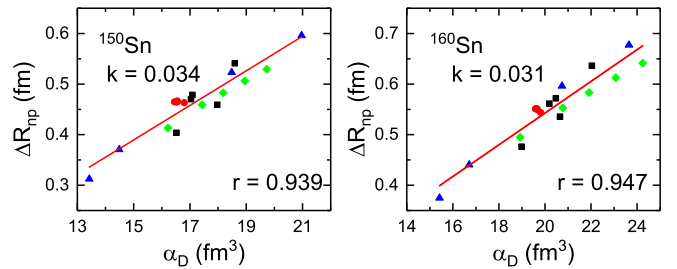}
	\caption{(Color online) Plots for neutron-skin thickness $\Delta R_\textnormal{np}$ against dipole polarizability $\alpha_D$ in $^{150,160}$Sn calculated by QRPA based on HFB with 19 Skyrme density functionals:  SIII, SIV, SV, SVI (blue up triangles); SLy230a, SLy230b, SLy4, SLy5, SLy8 (red circles);  SAMi, SAMi-J30, SAMi-J31, SAMi-J32, SAMi-J33 (green diamonds); SGI, SGII, SkM, SkM*, Ska (black squares). A regression line (red solid line) is obtained by a least-square linear fit of the calculated $\Delta R_\textnormal{np}$ as a function of $\alpha_D$. $r$ is Pearson's coefficient and $k$ (fm$^{-2}$) is the slope of regression line.} \label{fig-5}
\end{figure}

To understand the above strong linear correlations in neutron-rich Sn isotopes, we first investigate the role of pairing correlations. So in Fig. \ref{fig-2} the correlations between $\alpha_D$ and $L$ in $^{120,140,150,160}$Sn are studied without considering pairing effects. For stable nucleus $^{120}$Sn, the correlation between $\alpha_D$ and $L$ is similar as the case with pairing correlations, where the Pearson's coefficient is only slightly reduced without the inclusion of pairing correlations. However, for these three neutron-rich nuclei $^{140,150,160}$Sn, the linear correlations become much worse, where the Pearson's coefficients are largely reduced to the values $0.817$, $0.873$ and $0.867$, respectively, being all below $0.9$. It shows that the pairing correlations play important roles in keeping strong linear correlations between $\alpha_D$ and $L$ in neutron-rich Sn isotopes.

On the other hand, for neutron-rich nuclei, the PDR appears in the low-energy part of $E1$ transition strength distribution, which would give big contributions to the dipole polarizability. Since PDR  represents an oscillation between neutron skin and nearly isospin-saturated core, the correlations between its strengths and symmetry energy were also explored \cite{Piekarewicz_2006_PRC_PDR, Reinhard_2010_PRC_alphaD, Carbone_2010_PRC_PDR, Vretenar_2012_PRC_PDR}, although it is still an open question. Inspired by this, we extract the contributions of PDR to $\alpha_D$ in Sn isotopes in Fig. \ref{fig-3}, where the total dipole polarizabilities and contributions from PDR as functions of mass number $A$ in even-even Sn isotopes calculated by QRPA and RPA using Skyrme functional SLy4 are plotted. According to the dipole strength distributions and the transition densities,  different energies are selected as the upper boundaries of PDR for different Skyrme functionals, which are $9.0$ MeV for SVI, $10.0$ MeV for SIII, SLy family, SkM, SkM*, SGII, $10.5$ MeV for Ska, SAMi family, $11.0$ MeV for SGI, $12.0$ MeV for SIV, and $13.0$ MeV for SV.

Starting from $^{132}$Sn, the PDR appears and starts to contribute to the dipole polariziability $\alpha_D$. With the neutron number increases, the contribution from PDR becomes larger and larger, which dominates the evolution trend with mass number of the total $\alpha_D$. With the pairing correlations being turned off, the contribution from PDR to $\alpha_D$ is greatly reduced, which almost keeps a small constant with the increase of neutron number.  As a result, the total  $\alpha_D$ is also reduced a lot, and its increase trend with mass number becomes as slow as that before $^{132}$Sn. Before $^{132}$Sn, the pairing correlations only have very small influences on $\alpha_D$. Therefore, it can be seen that the pairing correlations play their important roles on dipole polariziabilities and further the linear correlations between $\alpha_D$ and $L$ through PDR.

In Fig. \ref{fig-4} we further study the correlation between dipole polarizabilities $\alpha_D$ contributed by PDR and the slope parameter $L$ of symmetry energy in  $^{134}$Sn, $^{140}$Sn, $^{150}$Sn, $^{160}$Sn isotopes. It shows that polarizability $\alpha_D$ of PDR has a good correlation with the slope parameter $L$ in general, which enhances the linear correlations between the total $\alpha_D$ and symmetry energy slope parameter $L$.

Apart from the correlation between $\alpha_D$ and $L$, the correlation between $\alpha_D$ and another important isovector property, i.e., neutron-skin thickness, is also investigated, and the plots for neutron-skin thickness against dipole polarizability in $^{150,160}$Sn are shown in Fig. \ref{fig-5}. Not surprisingly, the linear correlations between $\Delta R_{np}$ and $\alpha_D$ in $^{150}$Sn and $^{160}$Sn are strong with $r=0.939$ and $r=0.947$ respectively, since the neutron-skin thickness $\Delta R_{np}$ and $L$ are reported to have a good linear correlation when $|N-Z|$ is large \cite{B.A.Brown_2017_PRL_DRc}. The slopes $k$ of regression lines, fitted by $\Delta R_\textnormal{np}$ as a function of $\alpha_D$, are generally small in these neutron-rich nuclei, suggesting that $\alpha_D$ in neutron-rich nuclei can provide an effective constraints on neutron-skin thickness of the corresponding nuclei.

\subsection{$\alpha_D$ as a probe of nuclear isovector properties}\label{Sec-3.2}
\begin{table}[t]
	\caption{Constraints on the slope parameter of symmetry energy $L$ from experimental dipole polarizability values $\alpha_D^\textnormal{exp.}$ \cite{Tamii_2011_PRL_alphaD, Rossi_2013_PRL_alphaD,Birkhan_2017_PRL_alphaD, Bassauer_2020_PRC_dipole, Roca-Maza_2015_PRC_alphaD} using linear correlation between $\alpha_DJ$ and $L$ obtained by skyrme QRPA calculations using 19 Skyrme functionals. The Pearson's coefficient $r$ and the slope $k$ of the regression line fitted by  $\alpha_DJ$ as a function of $L$ are also given.  $J=31.7\pm3.2$MeV is adopted \cite{Oertel_2017_RMP_EoS}. $\Delta L_\textnormal{min}$ denotes the uncertainty coming from the uncertainty of $J$.}
	\renewcommand{\arraystretch}{1.3}\setlength{\tabcolsep}{0.3em}\setlength{\tabcolsep}{.3em} \label{tab-2}
	\centering
	{\small
	\begin{tabular}{cccccc} \hline \hline
		Nucleus      &  $\alpha_D^\textnormal{exp.}$ (fm$^3$) & $r $    & $k$ (fm$^3$)      &  $L$ (MeV)                   & $\Delta L_\textnormal{min}$ (MeV) \\ \hline
		$^{208}$Pb  &  $19.6\pm0.60$                         & 0.97    & 2.68     & $39.45\pm34.15$              &     $\pm 23.44$          \\  \hline
		$^{68}$Ni   &  $3.88\pm0.31$                         & 0.96    & 0.56     & $33.25\pm40.75$              &     $\pm 22.12$         \\  \hline
		$^{48}$Ca   &  $2.07\pm0.22$                         & 0.96    & 0.33     & $14.75\pm44.25$              &     $\pm 19.97$         \\  \hline
		$^{112}$Sn  &  $7.19\pm0.50$                         & 0.97    & 1.10     & $12.80\pm34.80$              &     $\pm 20.87$        \\  \hline
		$^{114}$Sn  &  $7.29\pm0.58$                         &  0.97   & 1.15     & $10.50\pm36.00$              &     $\pm 20.22$         \\  \hline
		$^{116}$Sn  &  $7.52\pm0.51$                         & 0.97    & 1.23     & $12.25\pm32.75$              &     $\pm 19.52$       \\  \hline
		$^{118}$Sn  &  $7.91\pm0.87$                         & 0.97    & 1.32     & $18.75\pm40.75$              &     $\pm 19.24$       \\  \hline
		$^{120}$Sn  &  $8.08\pm0.60$                         & 0.97    &  1.38    & $17.90\pm33.10$              &     $\pm 18.70$        \\  \hline
		$^{124}$Sn  &  $7.99\pm0.56$                         & 0.98    & 1.47     & \hskip0.5em$8.50\pm31.50$    &     $\pm 17.42$              \\  \hline \hline
	\end{tabular}}
\end{table}

\begin{figure}
	\centering
	\includegraphics[width=0.47\textwidth]{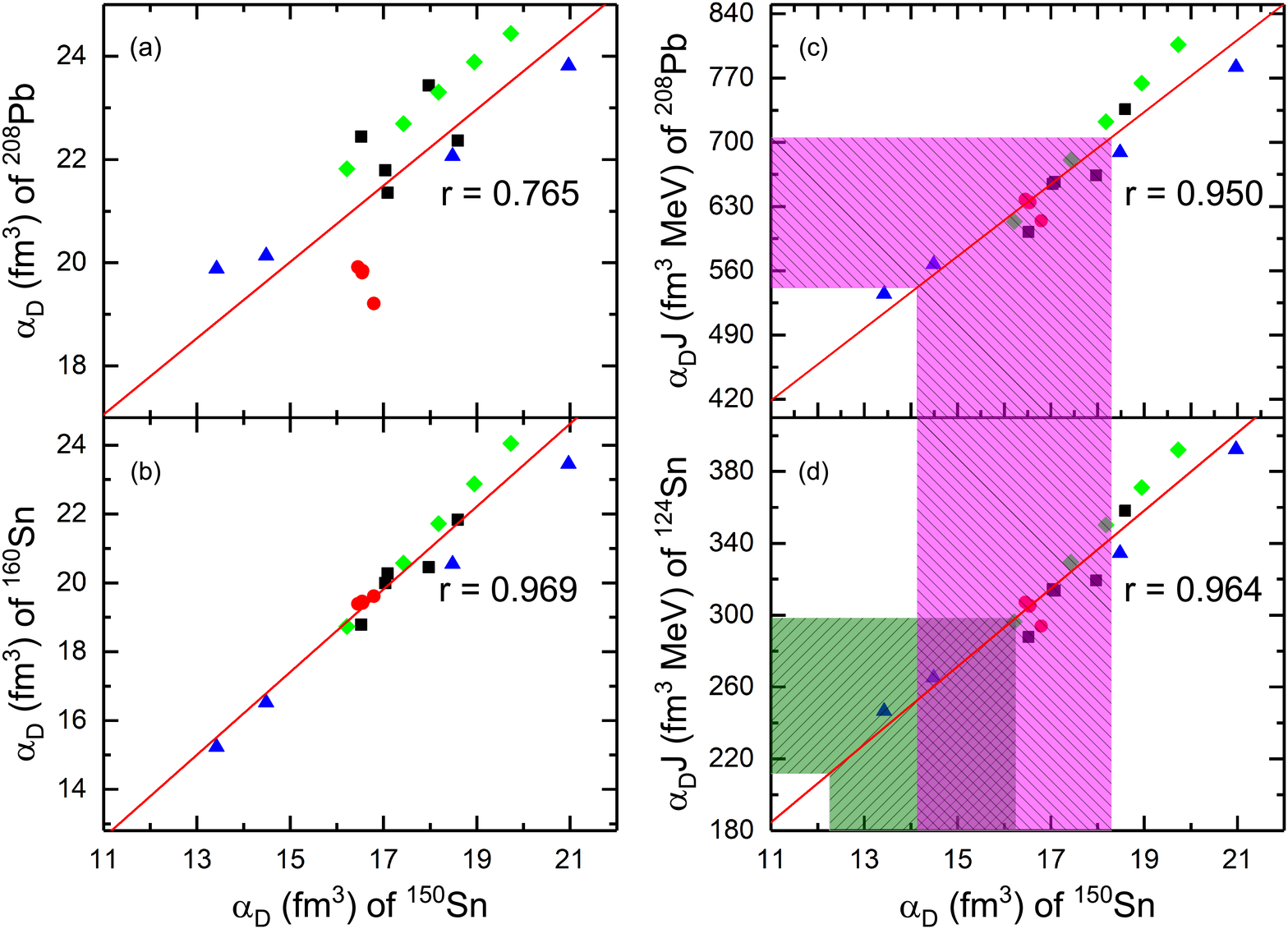}
	\caption{(Color online) The dipole polarizability $\alpha_D$ (a) in $^{208}$Pb and (b) in $^{160}$Sn as a function of the dipole polarizability in $^{150}$Sn.  The dipole polarizability $\alpha_D$ (c) in $^{208}$Pb and (d) in $^{124}$Sn times the symmetry energy at saturation density $J$ as a function of the dipole polarizability in $^{150}$Sn. Calculations are done by QRPA based on HFB with 19 Skyrme density functionals: SIII, SIV, SV, SVI (blue up triangles); SLy230a, SLy230b, SLy4, SLy5, SLy8 (red circles);  SAMi, SAMi-J30, SAMi-J31, SAMi-J32, SAMi-J33 (green diamonds); SGI, SGII, SkM, SkM*, Ska (black squares). $r$ is the Pearson's coefficient. Utilizing the experimental values of $\alpha_D$ in $^{208}$Pb \cite{Tamii_2011_PRL_alphaD, Roca-Maza_2015_PRC_alphaD} and in $^{124}$Sn \cite{Bassauer_2020_PRC_dipole}, and assuming $J=31.7\pm3.2$ MeV \cite{Oertel_2017_RMP_EoS}, the dipole polarizability of $^{150}$Sn is predicted to be between 14.13 and 16.25 fm$^3$.}  \label{fig-6}
\end{figure}

In Sec. \ref{Sec-3.1}, the correlations between $\alpha_D$ (or $\alpha_D J$) and nuclear isovector properties, e.g., $L$, $\Delta R_\textnormal{np}$, are investigated for the whole tin isotopes, so in the following, we will analyse what information we can obtain from these correlations, and which nucleus could be treated as a proper probe of nuclear isovector properties in terms of dipole polarizabilities.

Experimentally, the dipole polarizabilities of $^{208}$Pb \cite{Tamii_2011_PRL_alphaD}, $^{68}$Ni \cite{Rossi_2013_PRL_alphaD}, $^{48}$Ca \cite{Birkhan_2017_PRL_alphaD}, and stable Sn isotopes \cite{Bassauer_2020_PRC_dipole} were measured with high resolution. The correlations between $\alpha_DJ$ and $L$ are always strong for both stable nuclei and nuclei far from stability line from previous studies and our results in Sec. \ref{Sec-3.1}. So in Tab. \ref{tab-2}, we show the constraints on the slope parameter of symmetry energy $L$ from experimental dipole polarizability values $\alpha_D^\textnormal{exp.}$  using correlation between $\alpha_DJ$ and $L$ in these experimentally measured nuclei. The correlations between $\alpha_DJ$ and $L$ are obtained by QRPA calculations using 19 Skyrme density functionals as done in Sec.\ref{Sec-3.1}. The corresponding Pearson's coefficients $r$ and slopes $k$ of regression lines fitted by $\alpha_D J$ as a function of $L$ are also given in the table. It can be seen that the linear correlations are well kept for all these nuclei with $r>0.95 $. $J=31.7\pm3.2$ MeV from the statistic analysis of various available constraints \cite{Oertel_2017_RMP_EoS} is adopted for deducing the $L$ value. The uncertainty of $L$ is determined by $\Delta L = \big( J \Delta \alpha_D + \alpha_D \Delta J \big)/k$, where $\Delta \alpha_D$ and $\Delta J$ are the uncertainties of $ \alpha_D$ and $ J$, respectively. From Tab. \ref{tab-2}, it can be seen that $L$ have a remarkable uncertainties which are all larger than $\pm 30$ MeV. In the limiting case $\Delta \alpha_D=0$, the uncertainty of slope parameter $\Delta L_\textnormal{min}$ comes only from the the uncertainty of $J$, which is also given in Tab. \ref{tab-2}. It shows the uncertainty of $J$ contributes more than half of the total uncertainties of $L$, which hinders the effective constraints on $L$ from the correlation between $\alpha_D J$ and $L$. However, with the increase of neutron number in Sn isotopes, $\Delta L_\textnormal{min}$ has the tendency to become smaller. This is because the slope $k$ of regression line increases faster than the dipole polarizabity $\alpha_D$ with the increase of neutron number, and hence $\alpha_D/k$ becomes smaller. So the uncertainty caused by $\Delta J$ would become small if one finds a nucleus with a small $\alpha_D/k$ value.

Based on the analysis of neutron-rich Sn isotopes in Sec. \ref{Sec-3.1}, a strong correlation between $\alpha_D$ and $L$ appears in neutron-rich nuclei (seeing Fig. \ref{fig-1}) where the PDR gives a considerable contribution to the inverse energy-weighted sum rule $m_{-1}$. So it provides a more effective way to constrain $L$ directly from dipole polarizability. Moreover, the slope $k$ of regression line (in Fig. \ref{fig-1}) becomes larger with the increase of neutron number, which makes the constraints on $L$ from this correlation in neutron-rich nuclei more stiff. For example, an uncertainty of $\pm 0.5$ fm$^3$ in $\alpha_D$ of $^{140}$Sn, which is about the present accuracy for experimental measurement in dipole polarizability, could constrain $L$ within $\pm 10$ MeV, while with the same uncertainty of $\alpha_D$ in $^{160}$Sn,  $L$ can be constrained within $\pm 6$ MeV. However, for these neutron-rich nuclei, the experimental data for dipole polarizabilities is still unavailable, so we first need to make predictions on $\alpha_D$ in neutron-rich nuclei.
\begin{table}
	\caption{Predictions of the dipole polarizabilities in neutron-rich Sn isotopes from experimental dipole polarizabilities of $^{208}$Pb \cite{Tamii_2011_PRL_alphaD,Roca-Maza_2015_PRC_alphaD} and $^{124}$Sn \cite{ Bassauer_2020_PRC_dipole,Bassauer_2020_PLB_alphaD} using the correlations shown in Fig. \ref{fig-6} (c) and (d). The constrained values of slope parameter of symmetry energy $L$ and neutron-skin thickness of neutron-rich Sn isotopes are also given from the correlations shown in Figs. \ref{fig-1} and Figs. \ref{fig-5}. The Pearson's coefficients $r$ and slopes of regression line $k$ fitted by dipole polarizability $\alpha_D$ as a function of $L$, as well as by neutron-skin thickness $\Delta R_\textnormal{np}$ as a function of $\alpha_D$, are also shown respectively. }
	\renewcommand{\arraystretch}{1.3}\setlength{\tabcolsep}{0.3em}\setlength{\tabcolsep}{.24em} \label{tab-3}
	\centering
	{\footnotesize
		\begin{tabular}{cccccccc} \hline\hline
			\multirow{2}*{Nuclei}&  \multirow{2}*{$\alpha_D^{P}$ (fm$^3$)} &     \multicolumn{3}{c}{$\alpha_D$ as a function of $L$} &  \multicolumn{3}{c}{$\Delta R_\textnormal{np}$ as a function of $\alpha_D$} \\ \cline{3-8}
			           &     &  \hskip-0.8em$r$   & \hskip-1em\parbox{3em}{k{~\scriptsize(fm$^3$/MeV)}}       &  \hskip1.5em$L$ (MeV)  & r     & \hskip-1em\parbox{2em}{k~(fm$^{-2}$)}      &  $\Delta R_\textnormal{np}$(fm)   \\ \hline
			$^{140}$Sn &  $11.97\pm0.91$           &0.91    & 0.050             &  $18.5\pm18.1$ &   0.89  & 0.032  &  $0.295\pm0.029$               \\ \hline
			$^{142}$Sn &  $12.60\pm0.96$           &0.93    & 0.054             &  $19.4\pm17.7$ &   0.90  & 0.033  &  $0.316\pm0.031$              \\ \hline
			$^{144}$Sn &  $13.25\pm0.99$           &0.94    & 0.057             &  $20.3\pm17.3$ &   0.91  & 0.033  &  $0.338\pm0.033$              \\ \hline
			$^{146}$Sn &  $13.91\pm1.02$           &0.96    & 0.060             &  $21.1\pm16.9$ &   0.92  & 0.034  &  $0.358\pm0.034$              \\ \hline
			$^{148}$Sn &  $14.56\pm1.04$           &0.97    & 0.063             &  $21.7\pm16.5$ &   0.93  & 0.034  &  $0.377\pm0.035$             \\ \hline
			$^{150}$Sn &  $15.19\pm1.06$           &0.98    & 0.065             &  $22.3\pm16.2$ &   0.94  & 0.034  &  $0.396\pm0.036$             \\ \hline
			$^{152}$Sn &  $15.79\pm1.09$           &0.98    & 0.068             &  $22.7\pm16.0$ &   0.94  & 0.034  &  $0.414\pm0.037$             \\ \hline
			$^{154}$Sn &  $16.35\pm1.12$           &0.99    & 0.071             &  $23.1\pm15.7$ &   0.95  & 0.033  &  $0.431\pm0.038$             \\ \hline
			$^{156}$Sn &  $16.84\pm1.16$           &0.99    & 0.075             &  $23.5\pm15.5$ &   0.94  & 0.032  &  $0.447\pm0.038$              \\ \hline
			$^{158}$Sn &  $17.37\pm1.21$           &0.99    & 0.078             &  $23.5\pm15.5$ &   0.95  & 0.032  &  $0.461\pm0.039$              \\ \hline
			$^{160}$Sn &  $17.81\pm1.27$           &0.99    & 0.082             &  $23.5\pm15.5$ &   0.95  & 0.031  &  $0.474\pm0.040$             \\ \hline\hline
	\end{tabular}}
\end{table}

In Fig. \ref{fig-6}(a), we study the correlations of $\alpha_D$ between $^{208}$Pb and  $^{150}$Sn. Although it was found that $\alpha_D$ between two stable nuclei, e.g., between $^{208}$Pb and $^{120}$Sn, have a good linear correlation \cite{Hashimoto_2015_PRC_alphaD,Roca-Maza_2015_PRC_alphaD}, this correlation is no longer well kept when it is extended to $\alpha_D$ between one stable nucleus and one neutron-rich nucleus, e.g., between $^{208}$Pb and $^{150}$Sn, as seen in Fig. \ref{fig-6}(a). The correlation between two neutron-rich nuclei, e.g.,  between $^{160}$Sn and  $^{150}$Sn, is further checked in Fig. \ref{fig-6}(b), and it becomes strong again.  So one fails to predict  $\alpha_D$ of neutron-rich nuclei from $\alpha_D$ of stable nuclei directly.  Since both $\alpha_D J$ in stable nuclei and $\alpha_D$ in neutron-rich nuclei linearly correlate with $L$, $\alpha_D J$ in stable nuclei should also linearly correlate with $\alpha_D$ in neutron-rich nuclei. This is checked by our calculations in Fig. \ref{fig-6}, where $\alpha_D J$ in $^{208}$Pb (c) and in $^{124}$Sn (d) as a function of $\alpha_D$ in $^{150}$Sn are plotted.  Good linear correlations with $r=0.950$ and $0.964$ are found respectively, which can be used for the predictions of $\alpha_D$ in $^{150}$Sn as well as other neutron-rich nuclei. Utilizing the experimental  $\alpha_D$ values of $^{208}$Pb and $^{124}$Sn, shown in Tab. \ref{tab-2}, and adopting $J=31.7\pm3.2$ \cite{Oertel_2017_RMP_EoS}, $\alpha \in [12.26,16.25]$ fm$^3$ and $\alpha_D \in [14.13,18.29]$ fm$^3$ are obtained for $^{150}$Sn. The overlap $\alpha_D \in [14.13,16.25]$ fm$^3$ is finally taken as the predicted value for $^{150}$Sn.

The same process can be done for other neutron-rich nuclei. The predicted $\alpha_D$ from $^{140}$Sn to $^{160}$Sn are given in Tab. \ref{tab-3}, with which the corresponding constraints on $L$ and neutron-skin thickness $\Delta R_\textnormal{np}$  are deduced and presented in Tab. \ref{tab-3} from the correlations between $\alpha_D$ and $L$, as well as between $\Delta R_\textnormal{np}$ and $\alpha_D$. The corresponding Pearson's coefficients $r$ of both correlations are shown in the table, and it can be seen that the linear correlations are very well kept for all these neutron-rich nuclei.  Here since the $L$ values are constrained from the linear correlation between $\alpha_D$ and $L$ directly, the uncertainties become much smaller compared to those shown in Tab. \ref{tab-2}. With the increase of neutron number, the slope of regression line fitted by $\alpha_D$ as a function of $L$ becomes larger, and as a result, the uncertainty of $L$ also becomes smaller until $^{156}$Sn even with an increasing uncertainty in the predicted $\alpha_D^P$. For the neutron-skin thickness, the slope of regression line fitted by $\Delta R_\textnormal{np}$  as a function of $\alpha_D$ keeps almost a constant with increasing neutron numbers, yet the uncertainties of constrained neutron-skin thickness are becoming larger caused by the increasing uncertainties in $\alpha_D^P$. Due to the lack of experimental data of $\alpha_D$ in neutron-rich nuclei, the present constraints on $L$ shown in Tab. \ref{tab-3} in fact don't give new information compared to the $L$ values obtained from the correlation between $L$ and $\alpha_D J $ in $^{208}$Pb and in $^{124}$Sn. However, the direct correlation between $\alpha_D$ and $L$ would show its special importance and effectiveness in constraining nuclear isovector properties when the experimental data of $\alpha_D$ in neutron-rich tin isotopes are available, so the measurements of dipole polarizability towards neutron-rich nuclei are strongly called for.

\section{Summary}
The correlations between electric dipole polarizability $\alpha_D$ (or times symmetry energy at saturation density $J$) and slope parameter of symmetry energy $L$ are studied in Sn isotopes preformed by QRPA based on Skyrme HFB theory. The previously found correlation between $\alpha_DJ$ and $L$ is confirmed in the whole Sn isotopes from neutron-deficient ones to neutron-rich ones. The linear correlation between $\alpha_D$ and $L$ is not strong in stable tin isotopes and their surroundings, however, it becomes better for mass number $A>132$, and strong correlations are found when $A\ge140$ with the correlation coefficients $r>0.9$, where PDR gives a considerable contribution to $\alpha_D$. The enhancement of this correlation between $\alpha_D$ and $L$ is attributed to the pairing correlations, which play important roles through PDR.

With the available high-resolution data of $\alpha_D$, the constraints on $L$ are obtained from the correlation between $\alpha_DJ$ and $L$. Large uncertainties of $L$ are found, where more than half are contributed by the uncertainty from symmetry energy $\Delta J=\pm 3.2$ MeV. A proper candidate nucleus for constraining $L$ is the one with  a small  $\alpha_D/k$ value, where $k$ is the slope of regression line fitted by $\alpha_DJ$ as a function of $L$. In stable Sn isotopes, the $\alpha_D/k$ becomes smaller towards neutron-rich side.

With the strong correlation between $\alpha_D$ and $L$ in neutron-rich Sn isotopes, $L$ can be constrained directly and more stiffly if experimental data of $\alpha_D$ with high resolution in these nuclei are known. At the moment, $\alpha_D$ in neutron-rich nuclei are predicted using the linear correlation between  $\alpha_D J$ in a  stable nucleus with experimental data and  $\alpha_D$ in a neutron-rich nucleus. The measurements of electric dipole polarizability towards neutron-rich nuclei are called for.

\section*{Acknowledgement}
This work is partly supported by National Natural Science Foundation of China under Grant No. 12075104, 11675065 and 11875152, Fundamental Research Funds for the Central Universities under Grant No.lzujbky-2019-11, and Strategic Priority Research Program of Chinese Academy of Sciences, Grant No. XDB34000000.

\end{document}